\documentclass[aps,prd,showpacs,twocolumn]{revtex4}
\usepackage{eurosym}
\usepackage{graphicx}
\usepackage{amsmath}
\usepackage{bm}
\usepackage{color}
\usepackage{amssymb}

\def\be{\begin{equation}}
\def\ee{\end{equation}}
\def\bea{\begin{eqnarray}}
\def\eea{\end{eqnarray}}

\begin{document}

\title{Thermodynamic interpretation of the generalized gravity models with geometry - matter coupling}
\author{Tiberiu Harko}
\email{t.harko@ucl.ac.uk}
\affiliation{Department of Mathematics, University College London, Gower Street, London
WC1E 6BT, United Kingdom}

\begin{abstract}
Modified gravity theories with geometry - matter coupling, in which the action is an arbitrary function of the Ricci scalar and the matter Lagrangian  ($f\left(R,L_m\right)$ gravity), and of the Ricci scalar and of the trace of the matter energy-momentum tensor ($f(R,T)$ gravity), respectively, have the intriguing property that the divergence of the matter energy - momentum tensor is nonzero. In the present paper, by using the formalism of open thermodynamic systems, we interpret the generalized conservation equations in these gravitational theories from a thermodynamic point of view as describing irreversible matter creation processes, {\bf which could be validated by
fundamental particle physics}. Thus particle  creation corresponds to an irreversible energy flow from the gravitational field to the created matter constituents, with the second law of thermodynamics requiring that space-time transforms into matter. The equivalent particle number creation rates, the creation pressure and the entropy production rates are obtained for both $f\left(R,L_m\right)$ and $f(R,T)$ gravity theories. The temperature evolution laws of the newly created particles are also obtained. In the case of the $f(R,T)$ gravity theory the open irreversible thermodynamic interpretation of a simple cosmological model is presented in detail. It is also shown that due to the geometry--matter coupling, during the cosmological evolution a large amount of comoving entropy could be produced.
\end{abstract}

\date{\today }
\maketitle

\section{Introduction}

The recently  released Planck satellite data of the 2.7 degree Cosmic
Microwave Background  (CMB) full sky survey \cite{Planckresults}, as well as the measurement of the tensor modes
from large angle CMB B-mode polarisation by BICEP2 \cite{BICEP2} have generally
confirmed the standard $\Lambda $ Cold Dark Matter ($\Lambda $CDM)
cosmological model. The BICEP2 data, pointing towards a
tensor-to-scalar ratio $r = 0.2 ^{+0.07}_{-0.05}$, have provided a very
convincing evidence for the standard inflationary scenario, which predicts the generation of gravitational waves during the de
Sitter exponential expansion. However, there are some tensions between the BICEP2 results and the
Planck limits on standard inflationary models \cite{Mar}. On the other hand observations of the type Ia supernovae have convincingly shown that our universe presently experiences an accelerating evolution, for which the most natural explanation would be the presence of the dark energy, having an  equation of state parameter $w=-1.018\pm 0.057$ for a flat universe\cite{Bet}.

From a theoretical point of view, the necessity of explaining dark energy, as well as the second dominant component of the universe, dark matter \cite{Str} raises the fundamental question if the standard Einstein-Hilbert
action $S=\int{\left(R/2+L_m\right)\sqrt{-g}d^4x}$, where $R$ is the scalar
curvature, and $L_m$ is the matter Lagrangian density,  and in which matter is
minimally coupled to the geometry, can give an appropriate quantitative
description of the universe on all scales, going from the boundary
of the Solar System to the edge of the universe.

A theory with an explicit coupling between an arbitrary function of the
scalar curvature and the Lagrangian density of matter was proposed in
\cite{1}. The gravitational action of this gravity theory is of the form
$S=\int{\left\{f_1(R)+\left[1+\lambda
f_2(R)\right]L_m\right\}\sqrt{-g}d^4x}$. In these models an extra force
acting on massive test particles arises, and the motion is no longer
geodesic. Moreover, in this framework, one can also explain dark matter \cite{1}.
The initial ``linear''  geometry-matter coupling introduced in \cite{1} was
extended in \cite{2}, and a maximal extension of the Einstein-Hilbert
action with geometry-matter coupling, of the form $S=\int d^{4}x
\sqrt{-g}f\left(R,L_m\right)$ was considered in \cite{fL}. The cosmological and astrophysical implications of the $f\left(R,L_m\right)$ type gravity theories were investigated in \cite{litfLm}.

An alternative model to $f(R,L_m)$ gravity is the $f(R,T)$
gravity theory \cite{fT}, where $T$ is the trace of the matter energy-momentum tensor
$T_{\mu\nu}$. The corresponding action is given by $S=\int{\left[f\left(R,T\right)/2+
L_m\right]\sqrt{-g}d^4x}$. The dependence of the gravitational action on $T$ may be
due to the presence of  quantum effects (conformal anomaly), or of some
exotic imperfect fluids.  When the trace of the energy-momentum tensor
$T$  is zero, $T=0$, which is the case of the electromagnetic
radiation, the field equations of $f(R,T)$ theory reduce to those of
$f(R)$ gravity. The cosmological and astrophysical implications of the $f(R,T)$ gravity were considered in \cite{litfT}, and it was shown that such gravity theories explain the accelerating evolution of the universe. For a recent review of modified gravity theories with geometry-matter coupling see \cite{review}.

Both $f(R,L_m)$ and $f(R,T)$ gravity have the intriguing property that once geometry - matter coupling is introduced, the four - divergence of the energy - momentum tensor is non - zero. It is the purpose of the present paper to give a thermodynamic interpretation of the  gravity theories with geometry - matter coupling in the framework of the irreversible thermodynamic of open systems. More exactly, we consider that the non-conservativity of the matter energy - momentum tensor is related to irreversible matter creation processes, in which, due to the coupling between matter and geometry, there is an energy flow between the gravitational field and matter, with particles permanently added to the space -time.

The systematic investigation of the role of irreversible matter creation processes
 in general relativity and cosmology did
start with the papers \cite{Pri0} and \cite{Pri}. The description of
particle creation is based on the reinterpretation of the matter-energy
stress tensor in Einstein's equations, by modifying the usual adiabatic
energy conservation laws, and thereby including irreversible matter
creation. Thus matter creation corresponds to an irreversible energy flow
from the gravitational field to the created matter constituents. This point
of view results from consideration of the thermodynamics of open systems in
the framework of cosmology. In a gravitational system the second law of
thermodynamics requires that space-time transforms into matter, while the
inverse transformation is forbidden. As shown in \cite{Pri}, the usual
initial singularity associated with the Big Bang is structurally unstable
with respect to irreversible matter creation. The inclusion of dissipative
processes into the Einstein field equations lead to the possibility of
cosmological models that start from empty conditions and gradually build up
matter and entropy. Gravitational entropy takes a simple meaning as
associated to the entropy that is necessary to produce matter. This leads to
an extension of the third law of thermodynamics, as now the zero point of
entropy becomes the space-time structure out of which matter is generated
\cite{Pri0}.

The phenomenological approach for matter creation introduced in \cite%
{Pri0,Pri} was reformulated in a manifestly covariant way in \cite{Cal}. The
expressions for the entropy production rate and the temperature evolution
equation were also obtained, and some of their consequences were discussed.
The effects and implications of the irreversible matter creation processes on the cosmological evolution were extensively analyzed in \cite{Lima}-\cite{r5}. In particular, a theoretical model, called Creation of Cold Dark Matter (CCDM) was analyzed
in \cite{r}. It can be described macroscopically by introducing a negative pressure due
to matter creation. Therefore, the mechanism is capable to accelerate
the Universe, without the need of an additional dark energy component.   An alternative cosmological model, which assumes  the existence of gravitational
particle creation was studied in \cite{r0}.  The model fits well the supernova observations.  In this scenario one can alleviate the cosmic
coincidence problem, with dark matter and dark energy being of the same
nature, but acting at different scales.

Cosmological particle creation can  take place from the quantum
vacuum, due to external conditions,  which are caused by
the expansion or contraction of the Universe. This process was first discussed by Schr\"odinger \cite{q0}, who understood
that the cosmic evolution could lead to a mixing of positive and negative frequencies and that this would mean creation or destruction of matter due to the expansion of the Universe.  Later on this quantum phenomenon was studied via the modern techniques of
quantum field theory in curved space-times in \cite{q1} - \cite{q6}. In quantum field theory particle production is directly connected with the curvature of the Universe and, when the field equations are formulated in the form of harmonic oscillator equation without
friction part, the effect of gravity appears in the effective mass. The quantum rate of matter creation is maximum near the initial cosmological singularity, and post-inflationary re-heating may be explained from gravitational particle creation.

In this work, by using the formalism of the open irreversible thermodynamic systems, we reformulate the non - zero divergence of the energy - momentum tensor of the  ideal fluid in the $f\left(R,L_m\right)$ and $f(R,T)$ gravity theories  as an energy balance equation in the presence of matter creation. The particle number balance equation is also obtained, as well as the particle creation rate. The particle creation rate depends on both the thermodynamic parameters of the system, as well as of the functional form of the derivatives with respect to $L_m$ and $T$ of the function $f$ giving the geometry - matter coupling. When $f\equiv 0$, the system is conservative. The supplementary particle  creation pressure, induced by matter creation, is also determined, and it is shown that it is proportional to the particle creation rate. With the help of the creation pressure the divergence of the energy - momentum tensor can be rewritten in the form of an effective conservation equation, with the total pressure equal to the sum of the thermodynamic and creation pressure. The explicit form of the entropy, and of the entropy creation rates, associated to the particle transfer from geometry to matter, is also determined. The time evolution of the entropy is determined by the particle creation rates. The time evolution of the temperature of general thermodynamic systems with energy and pressure  depending on both particle numbers and temperature is also obtained. In the case of $f\left(R,L_m\right)$ gravity the behavior of the thermodynamic parameters describing matter creation are explicitly obtained for $L_m=-p$. The thermodynamic interpretation of a cosmological model corresponding to simple choice of the function $f(R,T)$ is also discussed in detail. Even that the approach proposed in the present paper is phenomenological,  quantum  particle creation processes \cite{q0}-\cite{q6} can give a microscopic justification to the proposed interpretation of the modified gravity models with geometry-matter coupling.

The present paper is organized as follows. In Section \ref{sect2} the field and conservation equations of the $f\left(R,L_m\right)$ and $f(R,T)$ gravity theories are briefly reviewed. The interpretation of the $f\left(R,L_m\right)$ gravity theory in the framework of the thermodynamic processes in open irreversible systems with matter creation is developed in Section \ref{sect3}. The irreversible thermodynamic interpretation of the $f(R,T)$ gravity theory is presented in Section~\ref{sect4}. We discuss and conclude our results in Section~\ref{sect5}. In the present paper we use the natural system of units with $c=8\pi G=1$, and the Landau-Lifshitz \cite{LaLi} conventions for the geometric and physical quantities.

\section{Generalized gravity models with geometry-matter coupling}\label{sect2}

In the present Section we briefly review two generalized gravity theories,
in which the four-divergence of the matter energy-momentum tensor is
non-zero. We consider the $f\left(R,L_m\right)$ and the $f(R,T)$ theories,
and we present the field and the conservation equations for each theory.

\subsection{$f\left(R,L_m\right)$ gravity}

The most general action $A$ for a $f\left(R,L_m\right)$ type modified theory of
gravity involving an arbitrary coupling between matter and curvature is
given by \cite{fL}
\begin{equation}
A=\int f\left(R,L_m\right) \sqrt{-g}\;d^{4}x~,
\end{equation}
where $f\left(R,L_m\right)$ is an arbitrary function of the Ricci scalar $R$%
, and of the Lagrangian density corresponding to matter, $L_{m}$. The only
requirement for the function $f\left(R,L_m\right)$ is to be an analytical
function of $R$ and $L_{m}$, respectively, that is, it must possess a Taylor
series expansion about any point. The matter energy-momentum tensor $T_{\mu
\nu}$ is defined as
\begin{equation}
T_{\mu \nu }=-\frac{2}{\sqrt{-g}}\frac{\delta \left( \sqrt{-g}L_{m}\right) }{%
\delta g^{\mu \nu }}.
\end{equation}

By assuming that the Lagrangian density $L_{m}$ of the matter depends only
on the metric tensor components, and not on its derivatives, we obtain $%
T_{\mu \nu }=L_{m}g_{\mu \nu }-2\partial L_{m}/\partial g^{\mu \nu }$.

Varying the action with respect to the metric tensor $g_{\mu \nu }$ we
obtain the field equations of the model as 
\begin{eqnarray}
f_{R}\left( R,L_{m}\right) R_{\mu \nu }+ \hat{P}_{\mu \nu } f_{R}\left(
R,L_{m}\right) -\frac{1}{2}\big[ f\left( R,L_{m}\right)  \notag \\
-f_{L_{m}}\left( R,L_{m}\right)L_{m}\big] g_{\mu \nu }= \frac{1}{2}%
f_{L_{m}}\left( R,L_{m}\right) T_{\mu \nu } ,  \label{feq}
\end{eqnarray}
where we have denoted $f_{R}\left( R,L_{m}\right) =\partial f\left(
R,L_{m}\right) /\partial R$ and $f_{L_m}\left( R,L_{m}\right) =\partial
f\left( R,L_{m}\right) /\partial L_{m}$, respectively, and we have
introduced the operator $\hat{P}_{\mu \nu }$, defined as
\begin{equation}
\hat{P}_{\mu \nu }=g_{\mu \nu }\square -\nabla _{\mu }\nabla _{\nu },
\end{equation}
with $\square =\nabla _{\mu }\nabla ^{\mu }$. The operator $\hat{P}_{\mu \nu
}$ has the property $\hat{P}_{\mu }^{ \mu }=3\square $.

By contracting the field equations Eq.~(\ref{feq}), we obtain the scalar
equation
\begin{equation}
3\square f_R +f_R R-2f =\left(\frac{1}{2}T-2L_m\right)f_{L_m} ,
\label{contr}
\end{equation}
where $T=T_{\mu }^{\mu }$ is the trace of the matter energy-momentum tensor.
By eliminating the term $\square f_R\left( R,L_{m}\right) $ between Eq.~(\ref%
{feq}) and Eq.~(\ref{contr}), we can reformulate the field equations as
\begin{eqnarray}
R_{\mu \nu }=\Lambda \left( R,L_{m}\right) g_{\mu \nu }+\frac{1}{f_R\left(
R,L_{m}\right) }\nabla _{\mu }\nabla _{\nu }f_R\left( R,L_{m}\right)  \notag
\\
+\Phi \left( R,L_{m}\right) \left( T_{\mu \nu }-\frac{1}{3}Tg_{\mu \nu
}\right) ,  \label{eqrmunu}
\end{eqnarray}
where we have denoted
\begin{equation}
\Lambda \left( R,L_{m}\right) =\frac{2f_R\left( R,L_{m}\right)R -f\left(
R,L_{m}\right) +f_{L_m}\left(R, L_{m}\right) L_{m}}{6f_R\left(
R,L_{m}\right) },  \label{lambda}
\end{equation}
and
\begin{equation}
\Phi \left( R,L_{m}\right) =\frac{f_{L_m}\left(R, L_{m}\right) }{f_R\left(
R,L_{m}\right) },
\end{equation}
respectively.

By taking the covariant divergence of Eq.~(\ref{feq}), with the use of the
mathematical identity \cite{Koi}
\begin{equation}
\left[\square, \nabla _{\nu}\right]F=\left(\square \nabla _{\nu }-\nabla
_{\nu}\square\right)F=R_{\mu \nu}\nabla ^{\mu }F,
\end{equation}
where $F$ is any arbitrary function of the space-time coordinates $x^{\mu}$, we obtain for the divergence of the energy-momentum tensor $T_{\mu \nu}$ the following relationship
\begin{eqnarray}
\nabla ^{\mu }T_{\mu \nu }&=&\nabla ^{\mu }\ln \left[ f_{L_m}\left(R,L_m%
\right)\right] \left( L_{m}g_{\mu \nu }-T_{\mu \nu }\right)  \notag \\
&=& 2\nabla ^{\mu }\ln \left[ f_{L_m}\left(R,L_m\right) \right] \frac{%
\partial L_{m}}{\partial g^{\mu \nu }}.  \label{noncons}
\end{eqnarray}

Generally the matter Lagrangian $L_{m}$ \ is a function of the matter energy
density $\rho $, the pressure $p$ as well as the other thermodynamic
quantities, such as the specific entropy $s$ or the baryon number $n$, so
that $L_{m}=L_{m}\left( \rho ,p,s,n\right) $. In the simple (but physically
the most relevant) case in which the matter obeys a barotropic equation of
state, so that the pressure is a function of the energy density of the
matter only, $p=p\left( \rho \right) $, the matter Lagrangian becomes a
function of the energy density only, and hence $L_{m}=L_{m}\left( \rho
\right) $. Then, the matter Lagrangian is given by \cite{L1,L2}
\begin{equation}  \label{lagr}
L_{m}\left( \rho \right) =\rho \left(1+\int_{0}^{p}\frac{dp}{\rho }%
\right)-p\left( \rho \right),
\end{equation}
while the energy-momentum tensor can be written as
\begin{equation}  \label{tens1}
T^{\mu \nu }=\left[ \rho +p\left( \rho \right) +\rho \tilde{\Pi} \left( \rho \right) %
\right] U^{\mu }U^{\nu }-p\left( \rho \right) g^{\mu \nu },
\end{equation}
respectively, where
\begin{equation}
\tilde{\Pi }\left( \rho \right) =\int_{0}^{\rho }\frac{p}{\rho ^{2}}d\rho
=\int_{0}^{p}\frac{dp}{\rho }-\frac{p\left(\rho \right)}{\rho }.
\end{equation}

The expression $\tilde{\Pi}(\rho) + p(\rho)/\rho$ represents the specific enthalpy
of the fluid. From a physical point of view $\tilde{\Pi }\left(\rho \right)$ can be
interpreted as the elastic (deformation) potential energy of the body, and
therefore Eq.~(\ref{tens1}) corresponds to the energy-momentum tensor of a
compressible elastic isotropic system.

\subsection{$f(R,T)$ gravity}

As a second example of a modified theory of gravity with non-conserved
energy - momentum tensor we consider the $f(R,T)$ gravity, with action
having the following form \cite{fT},
\begin{equation}  \label{fRT}
A=\frac{1}{16\pi}\int f\left(R,T\right)\sqrt{-g}\;d^{4}x+\int{L_m%
\sqrt{-g}\;d^{4}x} ,
\end{equation}
where $f\left(R,T\right)$ is an arbitrary function of the Ricci scalar, $R$,
and of the trace $T$ of the stress-energy tensor of the matter, $T_{\mu \nu}$%
. $L_m$ is the matter Lagrangian density. We define the variation
of $T$ with respect to the metric tensor as
\begin{equation}
\frac{\delta \left(g^{\alpha \beta }T_{\alpha \beta }\right)}{\delta g^{\mu
\nu}} =T_{\mu\nu}+\Theta _{\mu \nu} ,
\end{equation}
where
\begin{equation}
\Theta_{\mu \nu}\equiv g^{\alpha \beta }\frac{\delta T_{\alpha \beta }}{%
\delta g^{\mu \nu}}\, .
\end{equation}
By varying the gravitational action Eq.~(\ref{fRT}), we obtain the field
equations of the $f\left( R,T\right) $ gravity model as \cite{fT}
\begin{eqnarray}  \label{field}
&&f_{R}\left( R,T\right) R_{\mu \nu } - \frac{1}{2} f\left( R,T\right)
g_{\mu \nu } +\hat{P}_{\mu \nu} f_{R}\left( R,T\right) =  \notag \\
&&8\pi T_{\mu \nu}-f_{T}\left( R,T\right) T_{\mu \nu }-f_T\left(
R,T\right)\Theta _{\mu \nu} ,
\end{eqnarray}
where $f_R(R,T)=\partial f(R,T)/\partial R$ and $f_T(R,T)=\partial f(R,T)/\partial T$, respectively. Note that when $f(R,T)\equiv f(R)$, from Eqs.~(\ref{field}) we obtain the field equations of $f(R)$ gravity \cite{fR}.

By contracting Eq.~(\ref{field}) gives the following relation between the
Ricci scalar $R$ and the trace $T$ of the energy - momentum tensor,
\begin{eqnarray}\label{contr1}
&&f_{R}\left( R,T\right) R+3\square f_{R}\left( R,T\right) -2f\left(
R,T\right) =  \nonumber\\
&&8\pi T-f_{T}\left( R,T\right) T-f_{T}\left( R,T\right) \Theta ,
\label{contr}
\end{eqnarray}%
where we have denoted $\Theta =\Theta _{\mu }^{\ \mu }$.

By eliminating the term $\square f_{R}\left( R,T\right) $ between Eqs.~(\ref%
{field}) and (\ref{contr1}), the gravitational field equations can be
written in the form
\begin{eqnarray}
&&f_{R}\left( R,T\right) \left( R_{\mu \nu }-\frac{1}{3}Rg_{\mu \nu }\right)
+\frac{1}{6}f\left( R,T\right) g_{\mu \nu }=  \notag \\
&&8\pi \left( T_{\mu \nu }-\frac{1}{3}Tg_{\mu \nu }\right) -f_{T}\left(
R,T\right) \left( T_{\mu \nu }-\frac{1}{3}Tg_{\mu \nu }\right) -  \notag \\
&&f_{T}\left( R,T\right) \left( \Theta _{\mu \nu }-\frac{1}{3}\Theta g_{\mu
\nu }\right) +\nabla _{\mu }\nabla _{\nu }f_{R}\left( R,T\right) .
\end{eqnarray}%
By taking the covariant divergence of Eq.~(\ref{field}), with the
use of the following mathematical identity \cite{fT,Koi,Alv}
\bea
&&\nabla ^{\mu }\left[ f_{R}\left( R,T\right) R_{\mu \nu }-\frac{1}{2}f\left(
R,T\right) g_{\mu \nu }+\hat{P}_{\mu \nu} f_{R}\left( R,T\right) \right] \equiv \nonumber\\
&&\frac{1}{2}f_T(R,T)\nabla ^{\mu}T g_{\mu \nu},
\eea
 we obtain for the divergence of the energy - momentum  tensor $T_{\mu \nu }$ the equation
\bea\label{noncons1}
\nabla ^{\mu }T_{\mu \nu }&=&\frac{f_{T}\left( R,T\right) }{8\pi -f_{T}\left(
R,T\right) }\Bigg[ \left( T_{\mu \nu }+\Theta _{\mu \nu }\right) \nabla
^{\mu }\ln f_{T}\left( R,T\right) +\nonumber\\
&&\nabla ^{\mu }\Theta _{\mu \nu }-\frac{1}{2}\nabla ^{\mu}Tg_{\mu \nu}\Bigg].
\eea

For the tensor $\Theta _{\mu \nu }$ we find
\begin{equation}  \label{var}
\Theta _{\mu \nu}=-2T_{\mu \nu}+g_{\mu \nu }L_{m}-2g^{\alpha \beta }
\frac{\partial ^2L_{m}}{\partial g^{\mu \nu }\partial g^{\alpha \beta
}} .
\end{equation}

\section{Thermodynamic interpretation of the geometry-matter coupling in $%
f\left(R,L_m\right)$ gravity}\label{sect3}

In the present Section we investigate the $f\left(R,L_m\right)$ gravity
theory from the point of view of the thermodynamics of the matter creation
irreversible processes. As a first step in our study we obtain the energy
conservation equation, which, as compared to the standard adiabatic
conservation equation, contains an extra term, which can be interpreted as a
matter creation rate. Matter creation acts as an entropy source, and the
entropy flux and the temperature evolution of the gravitational system with
non - minimal geometry matter - coupling are obtained.

In the case of a perfect cosmological fluid with energy density $\epsilon$,
thermodynamic pressure $p$ and normalized four-velocity $U^\nu $ , $\nu
=0,1,2,3$, satisfying the condition $U_{\nu }U^{\nu }=1$, the energy -
momentum tensor $T_{\mu \nu}$ is given by
\begin{equation}  \label{em}
T_{\mu \nu }=\left(\epsilon +p\right)U_{\mu }U_{\nu }-pg_{\mu \nu}.
\end{equation}

We also introduce the projection operator $h_{\lambda }^{\nu}$, defined as
$h_{\lambda }^{\nu}=\delta _{\lambda }^{\nu}-U_{\lambda }U^{\nu}$,
with the property $U_{\nu}h^{\nu}_{\lambda }=0$.

\subsection{The energy conservation equation}

By taking the covariant divergence of Eq.~(\ref{em}) we obtain
\begin{eqnarray}
\nabla ^{\mu }T_{\mu \nu}&=&\left(\nabla ^{\mu }\epsilon +\nabla ^{\mu}
p\right)U_{\mu }U_{\nu }+\left(\epsilon +p\right)\nabla ^{\mu} U_{\mu}
U_{\nu}+  \notag \\
&&\left(\epsilon +p\right)U_{\mu}\nabla ^{\mu }U_{\nu}-\nabla ^{\mu }pg_{\mu
\nu}.
\end{eqnarray}
Therefore Eq.~(\ref{noncons}) takes the form
\begin{eqnarray}  \label{em1}
&&\left(\nabla ^{\mu }\epsilon +\nabla ^{\mu} p\right)U_{\mu }U_{\nu
}+\left(\epsilon +p\right)\nabla ^{\mu} U_{\mu} U_{\nu}+  \notag \\
&&\left(\epsilon +p\right)U_{\mu}\nabla ^{\mu }U_{\nu}-\nabla ^{\mu }pg_{\mu
\nu}=\nabla ^{\mu }\ln \left[ f_{L_m}\left(R,L_m\right)\right]\times  \notag
\\
&&\left[\left(L_m+p\right)g_{\mu \nu}-\left(\epsilon +p\right)U_{\mu}U_{\nu}%
\right].
\end{eqnarray}

By multiplying Eq.~(\ref{em1}) with $h_{\lambda }^{\nu}$ gives the momentum balance equation for a
perfect fluid  in the alternative $f\left(R,L_m\right)$ gravity theory as
\begin{eqnarray}  \label{force}
&&U^{\mu}\nabla _{\mu }U^{\lambda }=\frac{d^2x^{\lambda }}{ds^2}+\Gamma
_{\mu \nu}^{\lambda }U^{\mu }U^{\nu }=  \notag \\
&&\frac{1}{\epsilon +p}\left\{h^{\nu \lambda }\nabla _{\nu }p+h_{\mu
}^{\lambda }\nabla ^{\mu }\ln \left[ f_{L_m}\left(R,L_m\right)\right]%
\left(L_m+p\right)\right\}.  \notag \\
\end{eqnarray}

By multiplying Eq.~(\ref{em1}) with $U^{\nu}$, and by taking into account
the identity $U^{\nu}\nabla ^{\mu}U_{\nu}=0$ we finally obtain the energy
balance equation in the $f\left(R,L_m\right)$ gravity theory as
\begin{equation}  \label{em2}
\dot{\epsilon}+3\left(\epsilon +p\right)H=\frac{d}{ds}\ln \left[
f_{L_m}\left(R,L_m\right)\right]\left(L_m-\epsilon\right),
\end{equation}
where we have denoted $H=(1/3)\nabla ^{\mu }U_{\mu }$,
and $\dot{}=U^{\mu }\nabla _{\mu }=d/d{\tilde s}$, respectively, where $d{\tilde s}$ is the line element corresponding to the metric $g_{\mu \nu}$, $d{\tilde s}^2=g_{\mu \nu}dx^{\mu}dx^{\nu }$.

In the following we assume a homogeneous and isotropic cosmological model,
with line element given by the Friedmann-Robertson-Walker (FRW) metric,
\begin{equation}
d{\tilde s}^2=dt^2-a^2(t)\left(dx^2+dy^2+dz^2\right),
\end{equation}
where $a(t)$ is the scale factor. Moreover, we adopt a comoving coordinate
system with $U^{\mu }=(1,0,0,0)$. In the FRW geometry $H=\dot{a}/a$, $U^{\mu
}\nabla _{\mu }=\dot{}=d/dt$, and Eq.~(\ref{em2}) can be written in the equivalent form,
\begin{equation}  \label{20}
\frac{d}{dt}\left(\epsilon a^3\right)+p\frac{d}{dt}a^3=a^3\frac{d}{dt}\ln %
\left[ f_{L_m}\left(R,L_m\right)\right]\left(L_m-\epsilon\right).
\end{equation}

In the standard general relativistic case $f\left(R,L_m\right)=R/2+L_m$, which gives $f_{L_m}\left(R,L_m\right)=1$, with Eq.~(\ref{20}) reducing to the standard adiabatic matter conservation law $d\left(\epsilon a^3\right)+pda^3=0$. Moreover, due to our choice of the geometry, all the non-diagonal components of the energy--momentum tensor are equal to zero, so that $T_{\mu \nu}=0$, $\mu \neq \nu $. In particular this condition implies the impossibility of heat transfer in the FRW models of modified gravity with geometry--matter coupling, since the condition $T_{0i}\equiv 0$, $i=1,2,3$ must always hold.

\subsection{The matter and entropy creation rates}

In its general form, for a system containing $N$ particles in a volume $%
V=a^3 $, the second law of thermodynamics is given by \cite{Pri}
\begin{equation}  \label{21}
\frac{d}{dt}\left(\epsilon a^3\right)+p\frac{d}{dt}a^3=\frac{dQ}{dt}+\frac{h%
}{n}\frac{d}{dt}\left(na^3\right),
\end{equation}
where $dQ$ is the heat received by the system during time $dt$, $h=\epsilon
+p$ is the enthalpy per unit volume, and $n=N/V$ is the particle number
density. In the following we restrict our analysis to adiabatic transformations defined by the condition $dQ=0$, that is, we ignore proper heat transfer processes in the cosmological system. For these types of thermodynamic transformations  Eq.~(\ref{21}) represents
the formulation of the second law of thermodynamics that explicitly takes
into account the variation of the number of particles, described by the term
$(h/n)d\left(na^3\right)/dt$. Hence in the thermodynamic approach of open systems {\it the "heat" (internal energy) received by the system is
entirely due to the change in the number of particles}. From a cosmological
perspective, {\it this change is due to the transfer of energy from gravitation
to matter, with the creation of matter acting as a source of internal
energy}. Equivalently, for adiabatic transformations $dQ/dt=0$, Eq.~(\ref{21}%
) can be written as
\begin{equation}  \label{cons0}
\dot{\epsilon}+3(\epsilon +p)H=\frac{\epsilon +p}{n}\left(\dot{n}+3Hn\right).
\end{equation}

Therefore, from a thermodynamic point of view, Eq.~(\ref{20}), giving the
energy conservation equation in the $f\left(R,L_m\right)$ gravity theory,
can be interpreted as describing matter creation in an homogeneous and
isotropic geometry, with the particle number time variation given by the
equation
\begin{equation}  \label{22}
\dot{n}+3nH=\frac{n}{\epsilon +p}\frac{d}{dt}\ln \left[ f_{L_m}\left(R,L_m%
\right)\right]\left(L_m-\epsilon\right)=\Gamma n,
\end{equation}
where the particle creation rate $\Gamma $ is defined as
\begin{equation}\label{33}
\Gamma =\frac{1}{\epsilon +p}\frac{d}{dt}\ln \left[ f_{L_m}\left(R,L_m\right)%
\right]\left(L_m-\epsilon\right).
\end{equation}

Therefore the energy conservation equation can be written in the $%
f\left(R,L_m\right)$ gravity theory as
\begin{equation}  \label{41}
\dot{\epsilon}+3(\epsilon +p)H=(\epsilon +p)\Gamma.
\end{equation}

For adiabatic transformations Eq.~(\ref{21}), describing particle creation in an open systems, can be written as an effective energy conservation equation of the form
\be
\frac{d}{dt}\left(\epsilon a^3\right)+\left(p+p_c\right)\frac{d}{dt}a^3=0,
\ee
or, equivalently,
\be\label{comp}
\dot{\epsilon}+3\left(\epsilon +p+p_c\right)H=0,
\ee
where $p_c$, called the creation pressure, is defined as \cite{Pri}
\bea
p_c&=&-\frac{\epsilon +p}{n}\frac{d\left(na^3\right)}{da^3}=-\frac{\epsilon +p}{3nH}\left(\dot{n}+3nH\right)=\nonumber\\
&&-\frac{\epsilon +p}{3}\frac{\Gamma }{H}
\eea
Therefore in the $f\left(R,L_m\right)$ gravity theory the creation pressure is determined by the geometry - matter coupling, and is given by
\be\label{pc}
p_c=-\frac{1}{3H}\frac{d}{dt}\ln \left[ f_{L_m}\left(R,L_m\right)%
\right]\left(L_m-\epsilon\right).
\ee

The second law of thermodynamics can be written down by decomposing the
entropy change into an entropy flow $d_eS$ and an entropy creation $d_iS$,
so that the total entropy $S$ of the system is given by \cite{Pri0,Pri}
\begin{equation}
dS = d_eS + d_iS,
\end{equation}
with $d_iS > 0$. To evaluate the entropy flow and the entropy production, we
start from the total differential of the entropy \cite{Pri},
\begin{equation}
{\cal {T}}  d\left(sa^3\right)=d\left(\epsilon a^3\right)+pda^3-\mu d\left(na^3\right),
\end{equation}
where $\cal{T}$ is the temperature, $s=S/a^3$, and $\mu $ is the chemical
potential given by
\begin{equation}
\mu n=h-{\cal{T}}s.
\end{equation}

For closed systems and adiabatic transformations $dS=0$ and $d_iS=0$.
However, in the presence of matter creation there is a non-zero contribution
to the entropy. For homogeneous systems we still have $d_eS = 0$. In
contrast, matter creation contributes to the entropy production, and the
entropy time variation is given by \cite{Pri}
\begin{eqnarray}  \label{25}
{\cal{T}}\frac{d_iS}{dt}&=&{\cal{T}}\frac{dS}{dt}=\frac{h}{n}\frac{d}{dt}\left(na^3\right)-%
\mu \frac{d}{dt}\left(na^3\right)=  \notag \\
&&{\cal{T}}\frac{s}{n}\frac{d}{dt}\left(na^3\right)\geq 0.
\end{eqnarray}

From Eq.~(\ref{25}) we obtain for the time variation of the entropy the
equation
\begin{equation}\label{43}
\frac{dS}{dt}=\frac{S}{n}\left(\dot{n}+3Hn\right)\geq 0.
\end{equation}

With the use of Eq.~(\ref{22}), giving the particle number balance in $%
f\left( R,L_{m}\right) $ gravity theory, we obtain for the entropy
production the equation
\begin{equation}
\frac{1}{S}\frac{dS}{dt}=\frac{1}{\epsilon +p}\frac{d}{dt}\ln \left[
f_{L_{m}}\left( R,L_{m}\right) \right] \left( L_{m}-\epsilon \right) \geq 0.
\label{entfL}
\end{equation}%
The entropy flux vector $S^{\mu }$ is defined as \cite{Cal}
\begin{equation}
S^{\mu }=n\sigma U^{\mu },
\end{equation}%
where $\sigma =S/N$ is the specific entropy per particle. The entropy creation rate
must satisfy the second law of thermodynamics, $\nabla _{\mu }S^{\mu }\geq 0$%
. Since the specific entropy obeys the Gibbs relation \cite{Cal},
\begin{equation}
n{\cal{T}}d\sigma =d\epsilon -\frac{h}{n}dn,
\end{equation}%
and by taking into account the definition of the chemical potential $\mu $
of the system as
\begin{equation}
\mu =\frac{h}{n}-{\cal{T}}\sigma ,
\end{equation}%
we obtain
\bea\label{48}
\nabla _{\mu }S^{\mu }&=&\left( \dot{n}+3nH\right) \sigma +nU^{\mu}\nabla _{\mu }\sigma=\nonumber\\
&&\frac{1}{{\cal{T}}}\left(
\dot{n}+3Hn\right) \left( \frac{\epsilon +p}{n}-\mu \right) ,
\eea
where we have taken into account the relation
\bea
nT\dot{\sigma}=\dot{\epsilon}-\frac{\epsilon +p}{n}\dot{n}=0,
\eea
which follows immediately from Eq.~(\ref{cons0}).
With the use of Eq.~(\ref{22}) we obtain for the entropy production rate associated to
the particle creation processes in $f\left( R,L_{m}\right) $ gravity the
expression
\bea
\nabla _{\mu }S^{\mu }&=&\frac{1}{{\cal{T}}}\frac{n}{\epsilon +p}\frac{d}{dt}\ln \left[
f_{L_{m}}\left( R,L_{m}\right) \right] \times \nonumber\\
&&\left( L_{m}-\epsilon \right) \left(
\frac{\epsilon +p}{n}-\mu \right) .
\eea

In the general case of a perfect comoving fluid with two essential
thermodynamic variables, the particle number density $n$, and the
temperatures $\cal{T}$, it is conventional to express $\epsilon $ and $p$ in terms
of $n$ and $\cal{T}$ by means of the equilibrium equations of state,
\begin{equation}\label{51}
\epsilon =\epsilon (n,{\cal{T}}),p=p(n,{\cal{T}}).
\end{equation}
Then the energy conservation equation Eq.~(\ref{41}) takes the form
\begin{equation}
\frac{\partial \epsilon }{\partial n}\dot{n}+\frac{\partial \epsilon }{%
\partial {\cal{T}}}\dot{{\cal{T}}}+3(\epsilon +p)H=\Gamma n.
\end{equation}
With the help of the general thermodynamic relation \cite{Cal}
\begin{equation}
\frac{\partial \epsilon}{\partial n}=\frac{\epsilon +p}{n}-\frac{{\cal{T}}}{n}\frac{%
\partial p}{\partial {\cal{T}}},
\end{equation}
we obtain the temperature evolution of the newly created particles in the $%
f\left(R,L_m\right)$ gravity theory as
\begin{eqnarray}\label{54}
\frac{\dot{{\cal{T}}}}{{\cal{T}}}&=&\frac{\dot{n}}{n}\frac{\partial p}{\partial \epsilon }%
=\left(\Gamma -3H\right)\frac{\partial p}{\partial \epsilon}=  \notag \\
&&\frac{1}{\epsilon +p}\frac{\partial p}{\partial \epsilon}\Bigg\{ \frac{d}{%
dt}\ln \left[ f_{L_m}\left(R,L_m\right)\right]\left(L_m-\epsilon\right)-
\notag \\
&&3H(\epsilon +p)\Bigg\}.
\end{eqnarray}

\subsection{The case $L_m=-p$}

The exact form of the Lagrangian of the matter is one of the most intriguing
theoretical problems in general relativity. One possible choice is $L_{m}=-p$%
, which was used in \cite{La1,La2,La3} to derive the equation of motion of
test fluids in standard general relativity. In this case, as pointed out in
\cite{Fa}, the extra - force, given by Eq.~(\ref{force}), and which is one
of the distinguishing features of modified gravity theories with geometry -
matter coupling, identically vanishes. On the other hand, as argued in \cite%
{LoboL}, other choices for the matter Lagrangian, like $L_{m}=\epsilon $
or $L_{m}=-na$, where $a$ is the physical free energy defined
as $a=\epsilon /n - Ts$, are also possible \cite{La3,Haw}. In fact all
these expressions are on shell representations of a more general Lagrangian
density obtained through back-substitution of the equations of motion into
the related action \cite{La3}.

In the following we investigate the consequences of the choice $L_{m}=-p$
for the matter Lagrangian. In this case the extra-force in Eq.~(\ref{force})
vanishes, showing that in this model of $f\left( R,L_{m}\right) $ gravity
the motion of the test fluids is geodesic. On the other hand with this
choice the right hand of the energy conservation equation Eq.~(\ref{em2})
does not vanish, showing that the particle creation processes are present in
the model. The particle number balance equation Eq.~(\ref{22}) becomes
\begin{equation}
\dot{n}+3nH=-n\frac{d}{dt}\ln \left[ f_{-p}\left( R,-p\right) \right] ,
\end{equation}%
where $f_{-p}$ denotes the derivative of $f\left( R,-p\right) $ with respect
to $-p$, and can be immediately integrated to give
\begin{equation}
na^{3}=\frac{n_{0}a_{0}^{3}}{f_{-p}\left( R,-p\right) }.
\end{equation}

Therefore the variation of the total number of particles within a volume $V$
is inversely proportional to the derivative of $f$ with respect to the
matter Lagrangian. The energy conservation equation Eq. (\ref{41}) takes the
form
\begin{equation}
\dot{\epsilon}+3(\epsilon +p)H=-(\epsilon +p)\frac{d}{dt}\ln \left[
f_{-p}\left( R,-p\right) \right] .
\end{equation}

From Eq.~(\ref{pc}) the creation pressure $p_c$ is obtained as
\be
p_c=\frac{\epsilon +p}{3H}\frac{d}{dt}\ln \left[ f_{L_m}\left(R,L_m\right)%
\right].
\ee

For a radiation - like fluid with $p=\epsilon /3$ we obtain for the
variation of the energy density the expression
\begin{equation}
\epsilon a^{4}=\frac{\epsilon _{0}a_{0}^{4}}{\left[ f_{-p}\left( R,-p\right) %
\right] ^{4/3}},
\end{equation}%
where $\epsilon _{0}$ and $a_{0}$ are arbitrary constants of integration.
For the total entropy variation from Eq. (\ref{entfL}) we find
\begin{equation}
S=\frac{S_{0}}{f_{-p}\left( R,-p\right) },
\end{equation}%
where $S_{0}$ is an arbitrary constant of integration. In a cosmological
fluid where the density and pressure are functions of the temperature $T$
only, $\epsilon =\epsilon (T)$, $p=p(T)$, the entropy is given by \cite{Wein}
\begin{equation}
S=\frac{\left( \epsilon +p\right) V}{T}.
\end{equation}

For a radiation fluid $\epsilon =3p$, and it follows that the function $f$
satisfies the following partial differential equation,
\begin{equation}
\frac{\partial }{\partial \left( -p\right) }f\left( R,-p\right) =-\frac{%
S_{0}T(a)}{4\left( -p\right) a^{3}}.  \label{ented}
\end{equation}

By assuming a cosmological model with $f\left( R,-p\right)
=f_{0}(R(a))g_{0}\left( -p\right) $, Eq. (\ref{ented}) gives
\begin{equation}
f_{0}(R(a))=-\frac{S_{0}T(a)}{a^{3}},\frac{dg_{0}\left( -p\right) }{d\left(
-p\right) }=\frac{1}{4\left( -p\right) },
\end{equation}%
which fixes the matter Lagrangian dependence of the action as
\begin{equation}
g_{0}\left( L_{m}\right) =\frac{1}{4}\ln L_{m}.
\end{equation}

\section{Irreversible thermodynamic interpretation of the $f(R,T)$ gravity theory}\label{sect4}

In the present Section we consider the thermodynamic interpretation of the $f(R,T)$ gravity theory. In the present study we \textit{assume} again that the energy - momentum tensor of the matter is given by Eq.~(\ref{em}).  In order to obtain some explicit results
we fix the matter Lagrangian from the beginning as
\be
L_m=-p.
\ee

The trace of the matter energy - momentum tensor is given by $T=\epsilon -3p$. Then, with the use of Eq.~(%
\ref{var}), we obtain for the variation $\Theta _{\mu \nu}$ of the energy - momentum tensor of a perfect
fluid with respect to metric tensor $g_{\mu \nu}$  the expression \cite{fT}
\begin{equation}\label{theta1}
\Theta _{\mu \nu }=-2T_{\mu \nu }-pg_{\mu \nu }=-2(\epsilon +p)U_{\mu }U_{\nu}+pg_{\mu \nu}.
\end{equation}

\subsection{Energy and particle creation rates in $f(R,T)$ gravity}

With the use of Eq.~(\ref{theta1}) the divergence of the energy - momentum tensor in $f(R,T)$ gravity theory follows immediately from Eq.~(\ref{noncons1}) as
\bea\label{65}
\nabla ^{\mu }T_{\mu \nu}&=&-\frac{f_T(R,T)}{8\pi +f_T(R,T)}\Bigg[\left(\epsilon +p\right)U_{\nu}U_{\mu }\nabla ^{\mu }\ln f_T(R,T)+\nonumber\\
&&\nabla _{\nu}\frac{\epsilon -p}{2}\Bigg].
\eea
By multiplying Eq.~(\ref{65}) with $U^{\nu}$ gives
\bea\label{66}
U^{\nu}\nabla ^{\mu }T_{\mu \nu}&=&-\frac{f_T(R,T)}{8\pi +f_T(R,T)}\Bigg[\left(\epsilon +p\right)U_{\mu }\nabla ^{\mu }\ln f_T(R,T)+\nonumber\\
&&U^{\nu}\nabla _{\nu}\frac{\epsilon -p}{2}\Bigg].
\eea

From Eqs.~(\ref{66}), with the use of the explicit form of the matter energy - momentum tensor  we obtain the energy conservation equation in $f(R,T)$ gravity as
\bea\label{67}
&&\dot{\epsilon}+3(\epsilon +p)H=-\frac{f_T(R,T)}{8\pi +f_T(R,T)}\Bigg[\left(\epsilon +p\right)\times \nonumber\\
&&U_{\mu }\nabla ^{\mu }\ln f_T(R,T)+U^{\nu}\nabla _{\nu}\frac{\epsilon -p}{2}\Bigg].
\eea

We interpret again Eq.~(\ref{67}) as describing adiabatic irreversible thermodynamic particle creation in a cosmological context, according to Eq.~(\ref{cons0}). Therefore the particle balance equation is given by
\be
\dot{n}+3nH=\Gamma n,
\ee
where for the case of the $f(R,T)$ gravity the particle creation rate is defined as
\be\label{71}
\Gamma =-\frac{f_T(R,T)}{8\pi +f_T(R,T)}\Bigg[\frac{d}{dt}\ln f_T(R,T)+\frac{1}{2}\frac{\dot{\epsilon}-\dot{p}}{\epsilon +p}\Bigg].
\ee
The energy conservation equation can be formulated in an alternative way as
\be
\frac{d}{dt}\left(\epsilon a^3\right)+p\frac{d}{dt}a^3=\left(\epsilon +p\right)a^3\Gamma.
\ee

In the $f(R,T)$ gravity theory, the creation pressure corresponding to the particle production from the gravitational field, defined in Eq.~(\ref{pc}), is given by
\be\label{77}
p_c=\frac{\epsilon +p}{3H}\frac{f_T(R,T)}{8\pi +f_T(R,T)}\Bigg[\frac{d}{dt}\ln f_T(R,T)+\frac{1}{2}\frac{\dot{\epsilon}-\dot{p}}{\epsilon +p}\Bigg].
\ee

If $f(R,T)$ is independent of $T$ (the general relativistic limit), then $f_T(R,T)\equiv 0$, and both $\Gamma $ and $p_c$ vanish. Hence we reobtain the standard cosmological evolution without matter creation, and with the evolution of the Universe obeying a strict conservation law of the total energy.

From Eq.~(\ref{43}) we obtain the time variation of the entropy, which is entirely due to the matter creation processes,  as
\be
S(t)=S_0\exp\left[\int_0^t{\Gamma \left(t'\right)dt'}\right],
\ee
where $S_0$ is an arbitrary constant of integration. With the use of Eq.~(\ref{48}), we obtain for the entropy production rate the expression
\bea
\nabla _{\mu}S^{\mu}&=&\frac{n}{{\cal{T}}}\Gamma \left(\frac{\epsilon +p}{n}-\mu \right)=-\frac{n}{{\cal{T}}}\left(\frac{\epsilon +p}{n}-\mu \right)\times \nonumber\\
&& \frac{f_T(R,T)}{8\pi +f_T(R,T)}\Bigg[\frac{d}{dt}\ln f_T(R,T)+\frac{1}{2}\frac{\dot{\epsilon}-\dot{p}}{\epsilon +p}\Bigg].\nonumber\\
\eea

By assuming that the energy density and the thermodynamic pressure of the newly created particles are, according to Eqs.~(\ref{51}), functions of both the particle number $n$ and of the temperature $\cal{T}$, the temperature time variation of the created particles in the open thermodynamic system follows from Eq.~(\ref{54}), and in the $f(R,T)$ gravity is given by
\bea
\frac{1}{c_s^2}\frac{\dot{{\cal{T}}}}{{\cal{T}}}&=&\Gamma -3H=-\Bigg\{ \frac{f_T(R,T)}{8\pi +f_T(R,T)}\times \nonumber\\
&&\Bigg[\frac{d}{dt}\ln f_T(R,T)+\frac{1}{2}\frac{\dot{\epsilon}-\dot{p}}{\epsilon +p}\Bigg]+3H\Bigg\},
\eea
where $c_s^2=\partial p/\partial \epsilon $ is the speed of sound.

\subsection{Cosmological applications}

In the following we consider the thermodynamic interpretation of a simple $f(R,T)$ modified gravity theory \cite{fT}.  We assume
that the function $f(R,T)$ is given by
\be
f\left(R,T\right)=R+2g(T),
\ee
where $g(T)$ is an arbitrary function
of the trace of the energy - momentum tensor $T=\epsilon -3p$ of matter. The
gravitational field equations immediately follow from
Eq.~(\ref{field}), and are given by
\begin{equation}
R_{\mu\nu}-\frac{1}{2}Rg_{\mu\nu}
=8\pi T_{\mu \nu }-2g'\left(T\right)T_{\mu\nu}-2g'(T)
\Theta _{\mu \nu}+g(T)g_{\mu \nu }\, ,
\end{equation}
where the prime denotes a derivative with respect to the argument.

By taking the matter source as a perfect fluid, we have $\Theta _{\mu
\nu}=-2T_{\mu\nu}-pg_{\mu \nu}$, and then the field equations become
\begin{equation}
R_{\mu\nu}-\frac{1}{2}Rg_{\mu\nu}=8\pi T_{\mu\nu}
+2g'\left(T\right)T_{\mu\nu}+\left[2pg'(T)+g(T)\right]g_{\mu \nu } .
\end{equation}
By modeling the matter content of the universe as dust with $p=0$, the gravitational field equations
can be obtained as
\begin{equation}
R_{\mu\nu}-\frac{1}{2}Rg_{\mu\nu}
=8\pi T_{\mu\nu}+2g'(T)T_{\mu\nu}+g(T)g_{\mu\nu} .
\label{Ein1}
\end{equation}
These field equations were also proposed in \cite{Pop} to solve the cosmological constant problem. The simplest cosmological
model can be obtained by assuming a dust matter dominated  universe ($p=0$, $T=\rho
$), and by choosing the function $g(T)$ so that $g(T)=\lambda T$,
where $\lambda $ is a constant, giving.
\be
f_T(R,T)=2\lambda ={\rm constant}.
\ee

For the flat FRW metric, the
gravitational field equations are given by
\be
3H^2=\left(8\pi +3\lambda \right)\epsilon,
\ee
and
\be
2\dot{H}+3H^2=-\lambda \epsilon ,
\ee
respectively. Thus this $f(R,T)$ gravity model is equivalent to a
gravitational model with an effective cosmological constant
$\Lambda _\mathrm{eff}\propto H^2$ \cite{fT,Pop}. For this choice of $f(R,T)$ the gravitational
coupling becomes an effective time dependent coupling, of the
form $G_\mathrm{eff}=G\pm 2g'(T)$. Therefore the presence of the term $2g(T)$ in the
gravitational action modifies the nature of the gravitational interaction
between matter and geometry, replacing the gravitational constant $G$ by a running
gravitational coupling parameter.

The field equations reduce to a single equation for $H$,
\be
2\dot{H}+3\frac{8\pi +4\lambda }{8\pi +3\lambda }H^2=0 ,
\ee
with the general solution given by
\be H(t)=\frac{2\left(8\pi
+3\lambda \right)}{3\left(8\pi +4\lambda \right)}\frac{1}{t}=\frac{H_0}{t},
\ee
where $H_0=2\left(8\pi +4\lambda\right)/3\left(8\pi +3\lambda\right)$.
The scale factor evolves according to $a(t)=t^{H_0 }$, while the matter energy density of the universe can be obtained as
\be
\epsilon (t)=\frac{\epsilon _0}{t^2},
\ee
where $\epsilon _0=3H_0^2/\left(8\pi +3\lambda \right)$. The deceleration parameter $q$, defined as
\be
q=\frac{d}{dt}\frac{1}{H}-1,
\ee
is obtained as
\be
q=\frac{1}{H_0}-1=\frac{8\pi +\lambda }{2(8\pi +4\lambda )}.
\ee
If $q<0$, the expansion of the universe is accelerating, while $q>0$ corresponds to a decelerating phase.

The energy conservation equation follows from Eq.~(\ref{67}), and is given by
\be
\dot{\epsilon}+3\frac{8\pi +2\lambda }{8\pi +3\lambda }\epsilon H=0.
\ee
The particle creation rate, due to the energy transfer from the gravitational field to the matter, is obtained as
\be
\Gamma (t)=-\frac{\lambda }{8\pi +2\lambda }\frac{\dot{\epsilon}}{\epsilon}=\frac{3\lambda }{8\pi +3\lambda }H=\frac{\Gamma _0}{t},
\ee
where $\Gamma _0=\left(2\lambda /3\right)\left(8\pi+4\lambda \right)/\left(8\pi +3\lambda \right)^2$. The time variation of the particle number in the open matter - gravitation system is described by the equation
\be
\dot{n}=\frac{\Gamma _0-3H_0}{t}n,
\ee
and is given by
\be
n(t)=n_0t^{\Gamma _0-3H_0},
\ee
where $n_0$ is an arbitrary constant of integration. The creation pressure, obtained from Eq.~(\ref{77}), is determined as
\be
p_c(t)=\frac{\lambda }{3(8\pi +2\lambda )}\frac{\dot{\epsilon}}{H}=-\frac{\lambda }{8\pi +3\lambda }\epsilon =-\frac{p_{c0}}{t^2},
\ee
where $p_{c0}=\lambda \epsilon _0/(8\pi +3\lambda)$.
Finally, for the time variation of the entropy of the universe, which is due to the creation of the particles from the gravitational field, we obtain
\be
S(t)=S_0t^{\Gamma _0}.
\ee
Therefore we have obtained a full thermodynamic interpretation, in the framework of open irreversible thermodynamic systems, of this simple $f(R,T)$ cosmological model. The particle creation rate is proportional to the Hubble function, and it is a linearly decreasing function of time. The particle number created during this expansionary phase decreases during the cosmological evolution. The particle creation processes generate a large amount of comoving entropy, whose time dependence is given by a power law function.

\section{Discussions and final remarks}\label{sect5}

In the present paper we have investigated in a systematic way
the possibility that modified gravity theories with geometry - matter coupling  can be interpreted as providing a phenomenological
description of particle production in the cosmological fluid filling the
universe. One of the common features of these classes of gravity theories is the non-conservation of the matter energy -  momentum tensor. By using the mathematical formalism of the open thermodynamic systems we have provided a physical interpretation of the extra-terms generated by the non-minimal geometry  - matter coupling as describing particle production, with the gravitational field acting as a particle source. Both $f\left(R,L_m\right)$ and $f(R,T)$ gravity theories do admit such an interpretation, and the particle production rates, entropy, creation pressure and entropy generation rate have been explicitly obtained as functions of the coupling functions $f$, and of their derivatives, respectively.

The phenomenological cosmological particle creation formalisms considered in the present paper has been extensively discussed in
the literature within the context of standard general relativity.
Hence, the geometric coupling to matter of alternative gravity theories is not strictly necessary
for this type of cosmological approaches. However, there is a fundamental difference between the thermodynamic of open systems as formulated in the framework of standard general relativity, and in theories with geometry-matter coupling. While in general relativity the particle creation rates, and the creation pressure must be chosen either based on some plausible physical considerations, or  they must be inferred from some microscopic description, in both $f\left(R,L_m\right)$ and $f(R,T)$ modified gravity theories the coupling between geometry and matter completely determines the particle creation rates, the creation pressure and the entropy production, respectively. As one can see from Eqs.~(\ref{33}) and (\ref{pc}), in $f\left(R,L_m\right)$ gravity theory $\Gamma $ and $p_c$ is completely determined by the derivatives of the function $f\left(R,L_m\right)$ with respect to the matter Lagrangian $L_m$, and by the matter Lagrangian itself. In the case of $f(R,T)$ gravity theory, the particles creation rates and the creation pressure, given by Eqs.~(\ref{71}) and (\ref{77}), respectively, are also determined by the derivative of $f(R,T)$ with respect to $T$. Hence, in these classes of theories all the irreversible thermodynamic quantities are fully determined by the gravitational action.  In the general relativistic limit with $f\left(R,L_m\right)=R/2+L_m$, and $f(R,T)$ independent of $T$, all $\Gamma $ and $p_c$ are identically equal to zero. This result shows the essential role played by the coupling between matter and geometry in the phenomenological description of the cosmological particle production processes.

On the other hand, it has been suggested by Zeldovich \cite{Zeld} and later by
Murphy \cite{Mur} and \cite{Hu} that the introduction of viscosity in the cosmological
fluid is nothing but a phenomenological description of the effect of creation of
particles by the non-stationary gravitational field of the expanding universe. A
non-vanishing particle production rate is equivalent to a bulk viscous pressure in
the cosmological fluid, or, from a quantum point of view, with a viscosity of the
vacuum. This is due to the simple circumstance that any source term in the
energy balance of a relativistic fluid may be formally rewritten in terms of an
effective bulk viscosity.

The energy - momentum tensor of a relativistic fluid with bulk viscosity as the
only dissipative phenomena can be written as \cite{Mart}
\be
T_{\mu \nu}=\left(\epsilon +p+\Pi \right)U_{\mu}U_{\nu}-\left(p+\Pi\right),
\ee
where $\Pi $ is the bulk viscous pressure. The particle flow vector $N^{\mu}$ is given by $N^{\mu}=nU^{\mu}$. Limiting ourselves to second-order deviations from equilibrium, in the framework of causal thermodynamics the entropy
flow vector $S^{\mu}$ takes the form \cite{Isr}
\be
S^{\mu}=sN^{\mu}-\frac{\tau \Pi^2}{2\xi {\cal{T}}}U^{\mu},
\ee
where $\tau $ is the relaxation time, and $\xi $ is the coefficient of bulk viscosity. In the case of a homogeneous and isotropic geometry, the energy conservation equation in the presence of bulk viscous dissipative processes is obtained as
\be\label{comp1}
\dot{\epsilon}+3\left(\epsilon +p+\Pi\right)H=0.
\ee

By comparing Eq.~(\ref{comp1}), giving the conservation equation for a bulk viscous cosmological fluid, with Eq.~(\ref{comp}), which includes in the energy balance the creation of particles from the gravitational field, it follows that if one can take
\be
p_c=\Pi,
\ee
particle creation is equivalent with the introduction of an effective  bulk viscous pressure in the energy - momentum tensor of the cosmological fluid. Hence the causal bulk viscous pressure $\Pi $ acts as a creation pressure.   However, there is a major difference
between the particle creation irreversible processes in open thermodynamic systems, and bulk viscous processes, and this difference is related to the expression for entropy production rate. While the entropy production rate associated to particle creation is given by \cite{Cal}
\be
\nabla _{\mu }S^{\mu }=-\frac{3Hp_c}{{\cal{T}}}\left(1+\frac{\mu \Gamma n}{3Hp_c}\right)\geq 0,
\ee
in the presence of bulk viscous dissipative processes the entropy production rate can be obtained as \cite{Mart}
\be
\nabla _{\mu}S^{\mu}=-\frac{\Pi}{{\cal{T}}}\Bigg[3H+\frac{\tau }{\xi}\dot{\Pi}+\frac{\tau }{2\xi }\Pi\left(3H+\frac{\dot{\tau}}{\tau }-\frac{\dot {\xi}}{\xi}-\frac{\dot{{\cal{T}}}}{{\cal{T}}}\right)\Bigg].
\ee

In the particle creation model in open thermodynamics systems the entropy production rate is proportional to the creation pressure, while in the  viscous dissipative processes thermodynamic interpretation $\nabla _{\mu}S^{\mu}$ is quadratic in the creation pressure, $\nabla _{\mu}S^{\mu}\propto p_c^2/\xi{\cal{T}}$, and, moreover, involves a new dynamical variable, the bulk viscosity coefficient.

In the present paper we have adopted a phenomenological thermodynamical approach for the description of matter creation in cosmology, with the particle creation rate being determined by the coupling between matter and geometry. Such an approach cannot give any indication on the type or nature of the created particles. However, it is natural to expect that such creation processes are consistent with the similar processes that appear in the framework of the quantum particle creation in cosmology. Generally, a static gravitational field does not produce particles.  However, a time-dependent gravitational field generally produces particles.  In conventional quantum mechanics, tunneling from a small scale factor $a$ to large
$a$ would be described by the wave function which contains only outgoing waves as
$a\rightarrow \infty$ \cite{pap}. The decaying part of the wave function in the classically forbidden
region can be represented as $\Psi (a)\propto \exp\left\{-\int{\sqrt{2\left[V(a)-\epsilon\right]}da}\right\}$, where $V(a)$ is the quantum potential, and $\epsilon $ is the conformal energy of the scalar mode. When a massive scalar field is conformally coupled to gravity, with an action of the form $S_{\Phi}=\int{\left(g^{\mu \nu}\nabla _{\mu }\Phi \nabla _{\nu }\Phi /2-m_{\Phi}^2\Phi ^2+R\Phi ^2/12\right)\sqrt{-g}d^4x}$, unlike in conventional quantum mechanics, the resulting Schr\"{o}dinger  equation contains a negative kinetic term for the scale factor,  and the potential $V (a)$ also enters with a negative sign. Therefore in quantum cosmology, and quantum particle creation processes, the
excitation of the field $\Phi$  makes the tunneling processes exponentially easier, due to the increase of the effective value of $\epsilon$ \cite{pap}.   Therefore, the quanta $\Phi $ of the quantum field tend to be copiously created. Therefore, based on the quantum analogy, we expect that most of the particles created due to the geometry-matter coupling may be in the form of some scalar particles (bosons), that cosmologically may represent the dark matter content of the Universe. In the simple $f(R,T)$ cosmological model analyzed in the previous Section we have seen that the particle creation rate is maximum at the beginning of the cosmological expansion. Therefore we may also consider that most of the dark matter was created at the early stages of the cosmological evolution. On the other hand, self-interacting dark matter bosons may condense to form a Bose-Einstein condensate, and this hypothesis seems to be confirmed by the study of the galactic rotation curves \cite{bose1,bose2}.

One of the intriguing problems in modern cosmology is the entropy problem \cite{ent,ent1}. By using the open thermodynamic system interpretation, in the framework of theories with geometry - matter coupling models that start from empty conditions and gradually
build up matter and entropy can be easily constructed. In these models also the gravitational entropy
takes a simple meaning, being associated with the entropy that is necessary to
produce matter.  Matter creation process are modelled classically and on a
phenomenological level by means of a creation pressure, with particles
continuously added to the cosmological space-time. The dynamics and evolution of the  universe are
entirely determined by particle production processes, which, in turn, essentially depend on the geometry - matter coupling, and the influence of particle
production on geometry is essential.  As a
consequence of a large particle creation rate, inflationary behavior can also be obtained.
The evolution of the universe is determined, for a given equation of state, by the
numerical values of a single parameter $\Gamma $,  containing both the thermodynamics parameters of the system, as well as the geometry - matter coupling.  Matter creation processes
are naturally stopped after a finite interval of time, and the universe may end in a
decelerating era. A large amount of comoving entropy is produced during the
evolution of the universe.

The study of the irreversible matter--creation processes in the homogeneous and isotropic flat FRW geometry in models with geometry--matter coupling  opens the possibility of considering the viability of these gravity theories in
a cosmological context. However, in order to confirm the validity of the thermodynamic interpretation developed in the present paper, it is necessary to consider a wider range of cosmological and astrophysical tests of the $f\left(R,L_m\right)$ and $f(R,T)$ type theories. In particular, an essential test of these theories would be the analysis of their classical macroscopic
predictions in structure formation with linear perturbations (in the Newtonian
limit in small scales), and in observations fitting, including the study of the effects of matter creation on the CMB anisotropies.  Essentially the models introduced in the present paper are some simple toy models, whose main
interest is to begin probing and testing alternative gravity theories.

The gravitational theories considered in the present paper predict  the possibility that matter creation, associated
with geometry - matter coupling can also occur in the present - day universe, as initially considered
by Dirac \cite{22}. The existence of some forms of matter - geometry coupling
are not in contradiction with the cosmological observations or with some astrophysical data \cite{litfLm,litfT}, but observational evidence
of particle creation on a cosmological scale is still missing. Presumably, the functional forms of the functions $f$
entering in the present gravitational theories will be furnished by fundamental particle physics
models of decaying vacuum energy density and gravitational coupling, thus
permitting an in  depth comparison of the predictions of the theory with
observational data.

\end{document}